"Application of TensorFlow to recognition of visualized results of fragment molecular orbital (FMO) calculations"

Sona Saitou[1], Jun Iijima[1], Mayu Fujimoto[1], Yuji Mochizuki[1,2*], Koji Okuwaki[1], Hideo Doi[1], Yuto Komeiji[3#]

[1] *Department of Chemistry and Research Center for Smart Molecules, Faculty of Science, Rikkyo University, 3-34-1 Nishi-ikebukuro, Toshima-ku, Tokyo 171-8501, Japan*

[2] *Institute of Industrial Science, The University of Tokyo, 4-6-1 Komaba, Meguro-ku, Tokyo 153-8505, Japan*

[3] *Biomedical Research Institute, AIST, Tsukuba Central 2, Tsukuba, Ibaraki 305-8568, Japan*

\* Corresponding author: Yuji Mochizuki (fullmoon@rikkyo.ac.jp)

# Yuto Komeiji (y-komeiji@aist.go.jp)

(2018/1/24, JST - upload to arXiv)

Abstract

We have applied Google's TensorFlow deep learning toolkit to recognize the visualized results of the fragment molecular orbital (FMO) calculations. Typical protein structures of α-helix and β-sheet provide some characteristic patterns in the two-dimensional map of inter-fragment interaction energy termed as IFIE-map (Kurisaki et al., Biophys. Chem. 130 (2007) 1). A thousand of IFIE-map images with labels depending on the existences of α-helix and β-sheet were prepared by employing 18 proteins and 3 non-protein systems and were subjected to training by TensorFlow. Finally, TensorFlow was fed with new data to test its ability to recognize the structural patterns. We found that the characteristic structures in test IFIE-map images were judged successfully. Thus the ability of pattern recognition of IFIE-map by TensorFlow was proven.

# 1. Introduction

In recent years, the deep learning (DL) has attracted great interests and widely used as one of artificial intelligence technologies, especially for the recognitions and interpretations of speeches, patterns, and images [1,2]. Ref. [3] by Hinton et al. briefed the DL implementations (e.g., hidden layers and back propagations) and summarized typical applications. Goh et al. [4] provided a useful review of DL for the computational chemistry, indicating a promising potential of DL applications in that field. Several DL toolkits such as TensorFlow (Google) [5,6], Chainer (Preferred Networks) [7], Caffe (University of California, Berkeley) [8], and CNTK (Microsoft) [9] have been available, which have promoted the rapid extensions of DL. Currently, the most popular DL toolkit may be Google's TensorFlow [5,6] because of the amount of accessible examples and documents. Some applications of TensorFlow to chemistry have been reported [10-12]. For example, the molecular structures with molecular graphs were handled with TensorFlow for drug discovery in Ref. [10].

In this paper, we report a preliminary application of TensorFlow [5,6] to recognition of visualized results of the fragment molecular orbital (FMO) method [14-16], which has been the most widely used fragmentation-oriented scheme applicable to large molecules such as proteins [17]. It is well known [18,19] that the list of inter-fragment interaction energy (IFIE) [18,19] is informative in grasping the interactions among fragments of a given system. For example, the analyses with IFIE values have been extensively used in biophysical and pharmaceutical discussions of proteins with ligands [16]. The IFIE-map [20] is a two-dimensional visualization of IFIE results, by which various patterns of interaction can be understood at a glance (or without checking exhaustively a huge number of numerical values of IFIE) [21]. For protein systems, the typical structures of α-helix and β-sheet of proteins are identified with only characteristic patterns of stabilization and destabilization among amino acid residues in IFIE-map, where such a structure characterization is usually done with the Ramachandran plot of dihedral angle sequences in main chains [22]. Here, we attempt an automated structure identification only from the IFIE-map image data [20] by TensorFlow [5] without using the Ramachandran plot [22]. We try to exploit the power of image recognition with TensorFlow [5]. The remaining parts of this paper are composed as follows. Section 2 denotes the preparation of IFIE-maps images. The set-up of TensorFlow is described in Section 3, and the test results are presented in Section 4.

## 2. Preparation of IFIE-map

### 2.1. Proteins structures

At the training stage, a number of images (say a thousand) with variations should be necessary [3,4]. Since the purpose of the present study was an automated structure recognition of proteins from the IFIE-map image patterns [20], the three-dimensional structures of 18 different proteins were downloaded from the protein data bank (PDB [23]), as compiled in Table 1, where the existences of α-helix and β-sheet structures are indicated with True ("T") or False ("F") flags. The supervising labels for TensorFlow [5] were given as 0 for α-helix structure, 1 for β-sheet structure, and 2 for both structures. The PDB ID of the longest protein chain (in formal length of amino acid residues) was 5JAK.

Some conventional treatment such as omission of missing residues and addition of H-atoms was done for each protein in Table 1. The classical molecular dynamics (MD) simulation was performed to take the structural fluctuations into account under a hydrated condition. The AmberTools15 program suite [24] and the AMBER14SB FF parameters [25] were employed in MD. For each protein, 10 ns MD trajectory was generated at 300 K under the periodic boundary condition with an explicit treatment of water molecules and counter ions (Cl anion or Na cation). Other MD conditions were identical with those used in Ref. [26]. A single node server with an Intel Xeon processor was used for the parallelized MD runs. Dozens of sample structures (depending of proteins) were extracted from the MD trajectories with 1 ps interval, by which some variations should be introduced in IFIE-map images of each protein. The extracted cubic clusters were reshaped to the droplets with 4 Å water layer and counter ions for the subsequent FMO calculations [26]. As non-protein samples without both α-helix and β-sheet structures, three different systems of a peptoid (biomimetic molecules of proteins) [27], a water cluster (consisting of 64 molecules), and a pyridine aggregate (78 molecules) were prepared by MD simulations as well. The supervising label of non-proteins was set as 3.

*2.2. FMO calculations*

The prepared molecular structures of 18 proteins were subjected to the FMO calculations at the second-order Møller-Plesset perturbation (MP2) level [28-30] with the ABINIT-MP program [16] on several small in-house servers with Intel Xeon processors. The 6-31G basis set [31] was used to reduce the total cost of computations. We had experiences that the use of 6-31G set is acceptable for semi-quantitative discussions [32]. Since the number of IFIE-map images as the visualized FMO results [20] was sizable, we pursued this level of basis set to reduce the total cost of computations: the situation was different from usual DL studies with a rich amount of available image databases [33,34]. The standard fragmentation of residue-by-residue

[14,16] was adopted, and the water molecules and counter ions in the reshaped droplets [26] were treated as single fragments. Then, the IFIE results were converted to the IFIE-map images [20] on Windows PCs, by using the BioStation Viewer system [16], an associated graphical user interface of ABINIT-MP. For 18 proteins, only the residue-residue interactions were visualized in IFIE-map, and the interactions of water molecules and counter ions were omitted. About 50 IFIE-map images per protein were generated.

Here, an IFIE-map of Ubiquitin (1UBQ) is shown in the right side in Fig. 1 as an example. Each square image-unit in this IFIE-map illustrates the residue-by-residue IFIE value: 76 row by 76 column matrix. The upper and lower triangles of the matrix form of IFIE-map indicate the stabilization (as reddish square unit) and destabilization (as bluish square unit) between amino acid residues, respectively [20] (energy range from -50 kcal/mol to +50 kcal/mol).

As indicated in the left side of Fig. 1, Ubiquitin has both α-helix (reddish colored) and β-sheet (blueish colored). The red- and green-colored parts are visible in the horizontal and vertical bars associated with the IFIE-map matrix in the right side of Fig. 1, and they mean the corresponding α-helix and β-sheet partial structures: such guiding information is appended by the BioStation Viewer judged with the Ramachandran sequences of dihedral angles in the main chain [22]. For simplicity of description, the upper triangle of stabilization in IFIE-map [20] is focused on. The α-helix structure provides two characteristic patterns of square image-units as high density along the diagonal line and horizontal alignment. In contrast, the β-sheet structure shows the lines of square image-units with diagonal 45 degrees: both positive and negative directions are possible. When a given protein has both α-helix and β-sheet structures, for example Ubiquitin, the corresponding IFIE-map should show mixed features of both image characteristics.

Each IFIE-map image was processed by the InFanView tool [35] to extract only the matrix part consisting of square image-units. In other words, the two bars including the guide (or "answer") of structures (see again Fig. 1) were removed. Fig. 2 displays the extracted image examples of proteins having α-helix structure (5JRT), β-sheet structure (3K6D) and both structures (1UBQ). The total number of prepared IFIE-map images of proteins was 844, and that of non-proteins was 217, providing a total of 1061 images as a data stock for the training of TensorFlow [5].

### 3. Set-up of TensorFlow

TensorFlow version 0.8 [5] was installed on a PC (Intel Core i7) with Ubuntu 16.04

operating system. The number of hidden layers was initially two, where each layer consisted of a pair of the convolution layer and pooling layer (refer to Refs. [3-6] when needed). The activation function was of rectified linear unit (ReLU) type as a default setting of TensorFlow. Fig. 3 is a snapshot of the neural network (NN) structure of TensorFlow captured by the TensorBoard graphical interface [5]. The final normalized probabilities of judgement are output with the SoftMax function.

Under Python 2 environment, OpenCV (module name cv2) [36] and Numpy [37] were imported. OpenCV was utilized for image processing such as resizing without changing the original image data (from 28×28 pixels to 56×56 pixels in loading). For the convolution in NN, the settings of stride and padding were 1 and SAME, respectively. Randomly selected 20 images with the supervised labels (0, 1, 2, and 3) were used in each training step, where the pixel size of input images was varied in checking the dependence of the number of steps to achieve the judgment accuracy of 0.9. The cycle limit of training steps was set as 100. The cross entropy was employed as the loss function for the back propagation, and the Adam-gradient optimization was done (the learning rate being $1.0×10^{-4}$). The technique of dropout (up to 0.2) was introduced in both hidden layers (pool1 and pool2) and fully connected layer (fc1) to avoid the issue of overfitting. The kernel code for the final judgement is shown in Fig. 4. The trained data set was "model.ckpt". For each input image (28×28 pixels in loading) to be judged, the probabilities for the four labels were printed on the command line. Almost the same set-up was done in the case of three hidden layers as well: details omitted for simplicity.

4. Test results

   First, the effect of dropout is checked in the case of two layers. Fig. 5 plots the step counts which were required to achieve the judgement accuracy higher than 0.9, depending on the pixel sizes of loaded images. This measurement was iterated five times, and almost the same counts were observed. Blue lines are obtained by fixing the dropout of 0.2, whereas red lines are due to the cascade setting of dropout (0 for the 1 - 5 steps, 0.1 for 6 - 20 steps, and 0.2 for 21 - 100 steps). It is clear that the cascade setting of dropout reduces the cycle of training steps relative to the fixed one (about half). Larger image size should contain more detailed information of patterns, and thus the speed of training would be faster than that of smaller image size. When the pixel size grows, the count of required steps becomes smaller. Meanwhile, the convolution processing in NN formally scales as the square of pixel sizes: the actual timings of 20×20 pixels and 40×40 pixels per training step were consistently about 100 s and 400s, respectively, on a PC used for TensorFlow. Thus, certain compromise is necessary to lower the total cost of training, and then the pixel sizes ranging from 28×28 to 36×36 in image loading may be of plausible option (with the cascade setting of dropout).

Table 2 compiles the IFIE-map judgement results of 18 proteins by the two layers and three layers at the image loading by 28×28 pixels, where the sum of probabilities (for 0, 1, 2, and 3) are normalized to unity for each protein and where "T-Label" means the true label defined as the correct answer in Table 1. Note here that the test images of IFIE-map used for Table 2 were not used at the training stage. Similar to Table 2, the judgement results at the image loading by 32×32 pixels and 36×36 pixels are listed in Table 3 and Table 4, respectively. From Table 2, it is notable that the setting of two layers shows rather better performance than does that of three layers, in particular for the α-helix case. This result suggests a certain balance between the resolution at the image loading and the number of hidden layers. In fact, the performance of three layers setting becomes better for the case of 32×32 pixels, as shown in Table 3: in other words, more layers may require higher resolution [3,4]. At the image loading by 36×36 pixels in Table 4, the settings of both two and three layers provide the similar results with enough accuracy. In the results by the setting of three layers in Tables 2 and 3, it is observable that some proteins containing α-helix structure (with labels of 0 or 2) has relatively smaller probability values than do proteins having β-sheet only (label 1). This observation implies that the IFIE-map characteristics of β-sheet (expressed by square image-units) can be recognized more easily than those of α-helix. In conclusion, the present application of TensorFlow to the pattern recognition of IFIE-maps as the first trial has shown reasonable overall performance even for different sizes of protein. Visual inspection of the IFIE-maps by inexperienced human may not be competitive.

## 5. Summary

In this paper, we reported an application of Google's TensorFlow DL toolkit [5,6] to recognition of the IFIE-map images [20]. We have investigated whether α-helix and β-sheet structures can be identified without any information of dihedral angle [22] from the visualized FMO calculation results. A total of 18 proteins were used as samples (Table 1), where the sets of structures were generated by the classical MD simulations [26], and they were subjected to the FMO calculations with ABINIT-MP [16]. In addition to the proteins, three non-protein systems were also employed. Then, about a thousand of IFIE-map images were prepared through some automated operations, and they were input to TensorFlow with two or three hidden layers. After the training, a series of test images of proteins were successfully judged whether the structures of α-helix and β-sheet exist (Tables 2, 3, and 4). We would expect that computer-assisted interpretations just as DL [3,4] for the FMO results including structural fluctuations should become useful to handle such big data [38].

## Acknowledgements

This work was partly supported by Ministry of Education, Culture, Sports, Science and Technology (MEXT) as grant-in-aid (Kaken-hi) No. 16H04635 and also as a social and scientific priority issue #6 (Accelerated Development of Innovative Clean Energy Systems) to be tackled by using post-K computer (FS2020).

Table 1: PDB IDs [23] and structure labels of 18 proteins (see text).

| PDB ID | Length | α-helix | β-sheet | Label |
|--------|--------|---------|---------|-------|
| 1ACI   | 76     | T       | F       | 0     |
| 1UBQ   | 76     | T       | T       | 2     |
| 5JRT   | 77     | T       | F       | 0     |
| 1DCZ   | 77     | F       | T       | 1     |
| 1A91   | 79     | T       | F       | 0     |
| 5A7L   | 80     | T       | F       | 0     |
| 5G4D   | 88     | T       | T       | 2     |
| 1AB3   | 88     | T       | F       | 0     |
| 1A68   | 95     | T       | T       | 2     |
| 3K6D   | 99     | F       | T       | 1     |
| 5B0G   | 104    | T       | T       | 2     |
| 2NCO   | 102    | T       | F       | 0     |
| 5FRG   | 104    | T       | F       | 0     |
| 1CDB   | 105    | F       | T       | 1     |
| 1A57   | 116    | F       | T       | 1     |
| 5DRE   | 125    | T       | T       | 2     |
| 2NBG   | 125    | T       | T       | 2     |
| 5JAK   | 151    | T       | T       | 2     |

Table 2: Judgement results of 18 proteins (at image loading by 28×28 pixels).

| Seq. No. | PDB ID | 2 layers | | | | 3 layers | | | | T-Label |
|---|---|---|---|---|---|---|---|---|---|---|
| | | Prob. 0 | Prob. 1 | Prob. 2 | Prob. 3 | Prob. 0 | Prob. 1 | Prob. 2 | Prob. 3 | |
| 1 | 1ACI | 0.78 | 0.00 | 0.19 | 0.03 | 0.54 | 0.02 | 0.39 | 0.05 | 0 |
| 2 | 1UBQ | 0.00 | 0.00 | 1.00 | 0.00 | 0.01 | 0.00 | 0.99 | 0.00 | 2 |
| 3 | 5JRT | 0.99 | 0.00 | 0.00 | 0.01 | 0.73 | 0.08 | 0.06 | 0.13 | 0 |
| 4 | 1DCZ | 0.00 | 1.00 | 0.00 | 0.00 | 0.02 | 0.82 | 0.14 | 0.01 | 1 |
| 5 | 1A91 | 0.94 | 0.00 | 0.01 | 0.05 | 0.51 | 0.00 | 0.32 | 0.16 | 0 |
| 6 | 5A7L | 0.98 | 0.00 | 0.00 | 0.01 | 0.78 | 0.01 | 0.12 | 0.09 | 0 |
| 7 | 5G4D | 0.00 | 0.00 | 1.00 | 0.00 | 0.06 | 0.13 | 0.79 | 0.02 | 2 |
| 8 | 1AB3 | 0.96 | 0.00 | 0.01 | 0.03 | 0.64 | 0.00 | 0.23 | 0.13 | 0 |
| 9 | 1A68 | 0.00 | 0.00 | 0.99 | 0.01 | 0.24 | 0.01 | 0.73 | 0.03 | 2 |
| 10 | 3K6D | 0.00 | 1.00 | 0.00 | 0.00 | 0.01 | 0.98 | 0.00 | 0.01 | 1 |
| 11 | 5B0G | 0.00 | 0.00 | 1.00 | 0.00 | 0.03 | 0.02 | 0.94 | 0.01 | 2 |
| 12 | 2NCO | 1.00 | 0.00 | 0.00 | 0.00 | 0.99 | 0.00 | 0.00 | 0.00 | 0 |
| 13 | 5FRG | 1.00 | 0.00 | 0.00 | 0.00 | 0.99 | 0.00 | 0.01 | 0.00 | 0 |
| 14 | 1CDB | 0.00 | 1.00 | 0.00 | 0.00 | 0.02 | 0.89 | 0.09 | 0.00 | 1 |
| 15 | 1A57 | 0.00 | 1.00 | 0.00 | 0.00 | 0.02 | 0.93 | 0.01 | 0.05 | 1 |
| 16 | 5DRE | 0.00 | 0.00 | 0.99 | 0.01 | 0.19 | 0.06 | 0.63 | 0.13 | 2 |
| 17 | 2NBG | 0.00 | 0.00 | 1.00 | 0.00 | 0.04 | 0.00 | 0.96 | 0.00 | 2 |
| 18 | 5JAK | 0.03 | 0.00 | 0.94 | 0.03 | 0.27 | 0.00 | 0.70 | 0.03 | 2 |

Table 3: Judgement results of 18 proteins (at image loading by 32×32 pixels).

| Seq. No. | PDB ID | 2 layers | | | | 3 layers | | | | T-Label |
|---|---|---|---|---|---|---|---|---|---|---|
| | | Prob. 0 | Prob. 1 | Prob. 2 | Prob. 3 | Prob. 0 | Prob. 1 | Prob. 2 | Prob. 3 | |
| 1 | 1ACI | 0.74 | 0.00 | 0.26 | 0.00 | 0.74 | 0.01 | 0.19 | 0.06 | 0 |
| 2 | 1UBQ | 0.00 | 0.00 | 1.00 | 0.00 | 0.02 | 0.00 | 0.98 | 0.00 | 2 |
| 3 | 5JRT | 1.00 | 0.00 | 0.00 | 0.00 | 0.88 | 0.03 | 0.02 | 0.06 | 0 |
| 4 | 1DCZ | 0.00 | 1.00 | 0.00 | 0.00 | 0.08 | 0.91 | 0.01 | 0.00 | 1 |
| 5 | 1A91 | 0.99 | 0.00 | 0.00 | 0.01 | 0.68 | 0.01 | 0.02 | 0.30 | 0 |
| 6 | 5A7L | 0.98 | 0.00 | 0.01 | 0.01 | 0.67 | 0.02 | 0.03 | 0.28 | 0 |
| 7 | 5G4D | 0.00 | 0.00 | 1.00 | 0.00 | 0.15 | 0.03 | 0.77 | 0.05 | 2 |
| 8 | 1AB3 | 0.99 | 0.00 | 0.00 | 0.00 | 0.82 | 0.01 | 0.01 | 0.16 | 0 |
| 9 | 1A68 | 0.01 | 0.00 | 0.99 | 0.00 | 0.19 | 0.01 | 0.78 | 0.02 | 2 |
| 10 | 3K6D | 0.00 | 1.00 | 0.00 | 0.00 | 0.08 | 0.82 | 0.07 | 0.04 | 1 |
| 11 | 5B0G | 0.01 | 0.00 | 0.99 | 0.00 | 0.13 | 0.11 | 0.74 | 0.02 | 2 |
| 12 | 2NCO | 1.00 | 0.00 | 0.00 | 0.00 | 0.90 | 0.05 | 0.04 | 0.01 | 0 |
| 13 | 5FRG | 0.99 | 0.00 | 0.00 | 0.01 | 0.83 | 0.04 | 0.03 | 0.10 | 0 |
| 14 | 1CDB | 0.00 | 1.00 | 0.00 | 0.00 | 0.02 | 0.96 | 0.01 | 0.01 | 1 |
| 15 | 1A57 | 0.00 | 0.99 | 0.00 | 0.00 | 0.12 | 0.82 | 0.03 | 0.03 | 1 |
| 16 | 5DRE | 0.04 | 0.05 | 0.90 | 0.01 | 0.18 | 0.24 | 0.56 | 0.02 | 2 |
| 17 | 2NBG | 0.00 | 0.00 | 0.99 | 0.00 | 0.06 | 0.03 | 0.90 | 0.00 | 2 |
| 18 | 5JAK | 0.01 | 0.00 | 0.99 | 0.00 | 0.05 | 0.04 | 0.91 | 0.00 | 2 |

Table 4: Judgement results of 18 proteins (at image loading by 36×36 pixels).

| Seq. No. | PDB ID | 2 layers | | | | 3 layers | | | | T-Label |
|---|---|---|---|---|---|---|---|---|---|---|
| | | Prob. 0 | Prob. 1 | Prob. 2 | Prob. 3 | Prob. 0 | Prob. 1 | Prob. 2 | Prob. 3 | |
| 1 | 1ACI | 0.87 | 0.00 | 0.12 | 0.00 | 0.87 | 0.00 | 0.13 | 0.00 | 0 |
| 2 | 1UBQ | 0.00 | 0.00 | 1.00 | 0.00 | 0.00 | 0.00 | 1.00 | 0.00 | 2 |
| 3 | 5JRT | 1.00 | 0.00 | 0.00 | 0.00 | 1.00 | 0.00 | 0.00 | 0.00 | 0 |
| 4 | 1DCZ | 0.00 | 1.00 | 0.00 | 0.00 | 0.00 | 1.00 | 0.00 | 0.00 | 1 |
| 5 | 1A91 | 0.99 | 0.00 | 0.01 | 0.00 | 1.00 | 0.00 | 0.00 | 0.00 | 0 |
| 6 | 5A7L | 1.00 | 0.00 | 0.00 | 0.00 | 1.00 | 0.00 | 0.00 | 0.00 | 0 |
| 7 | 5G4D | 0.00 | 0.00 | 1.00 | 0.00 | 0.00 | 0.00 | 1.00 | 0.00 | 2 |
| 8 | 1AB3 | 1.00 | 0.00 | 0.00 | 0.00 | 1.00 | 0.00 | 0.00 | 0.00 | 0 |
| 9 | 1A68 | 0.00 | 0.00 | 1.00 | 0.00 | 0.00 | 0.00 | 1.00 | 0.00 | 2 |
| 10 | 3K6D | 0.00 | 1.00 | 0.00 | 0.00 | 0.00 | 1.00 | 0.00 | 0.00 | 1 |
| 11 | 5B0G | 0.00 | 0.00 | 1.00 | 0.00 | 0.00 | 0.00 | 1.00 | 0.00 | 2 |
| 12 | 2NCO | 1.00 | 0.00 | 0.00 | 0.00 | 1.00 | 0.00 | 0.00 | 0.00 | 0 |
| 13 | 5FRG | 1.00 | 0.00 | 0.00 | 0.00 | 1.00 | 0.00 | 0.00 | 0.00 | 0 |
| 14 | 1CDB | 0.00 | 1.00 | 0.00 | 0.00 | 0.00 | 1.00 | 0.00 | 0.00 | 1 |
| 15 | 1A57 | 0.00 | 1.00 | 0.00 | 0.00 | 0.00 | 1.00 | 0.00 | 0.00 | 1 |
| 16 | 5DRE | 0.00 | 0.00 | 0.99 | 0.01 | 0.00 | 0.00 | 0.99 | 0.00 | 2 |
| 17 | 2NBG | 0.00 | 0.00 | 1.00 | 0.00 | 0.00 | 0.00 | 1.00 | 0.00 | 2 |
| 18 | 5JAK | 0.00 | 0.00 | 0.99 | 0.00 | 0.00 | 0.00 | 1.00 | 0.00 | 2 |

Fig. 1: Left; Structure of Ubiqutin (1UBQ) with α-helix (red) and β-sheet (blue) presentation. Right; IFIE-map of 1UBQ for residue-residue interactions (see text).

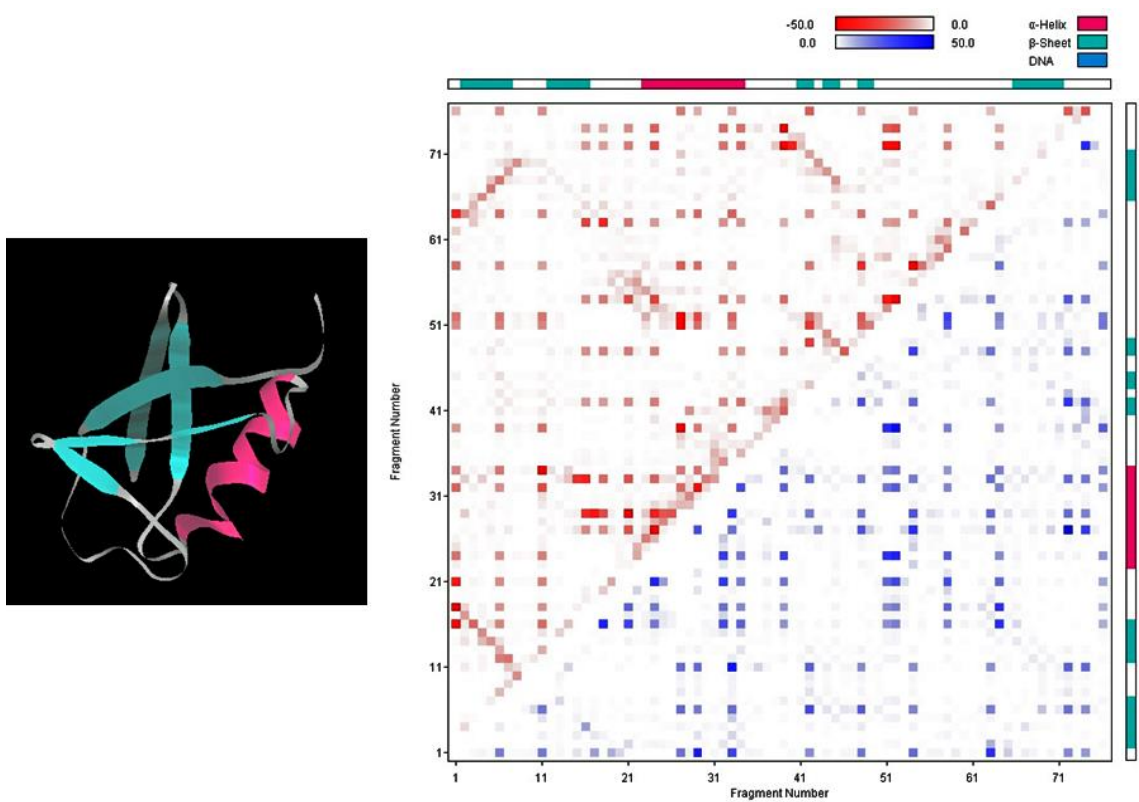

Fig. 2: Typical images of IFIE-map (extracted matrix part). 5JRT (Left), 3K6D (Center), and 1UBQ (Right) are used as examples with α-helix structure (label 0), β-sheet structure (label 1), and both structures (label 2).

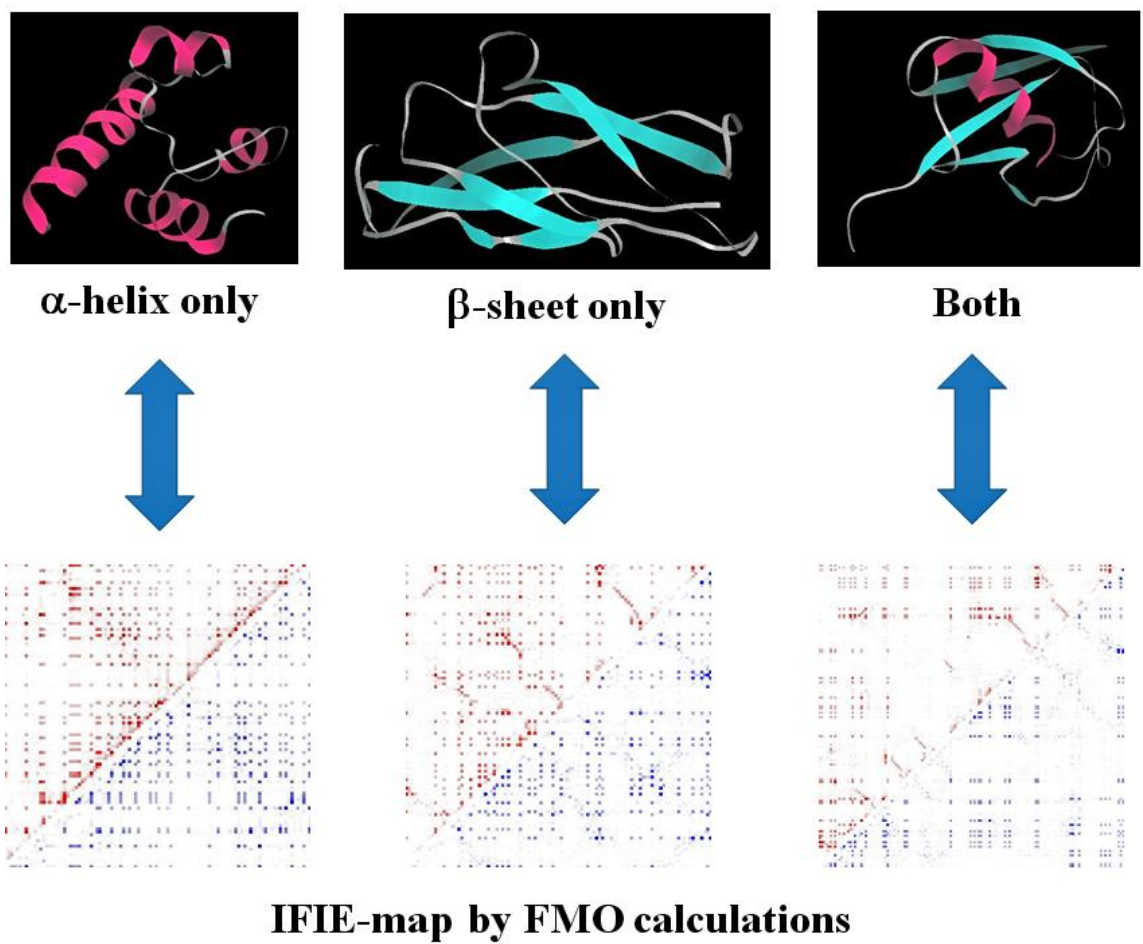

Fig. 3: NN structure of TensorFlow with two hidden layers (captured by TensorBoard).

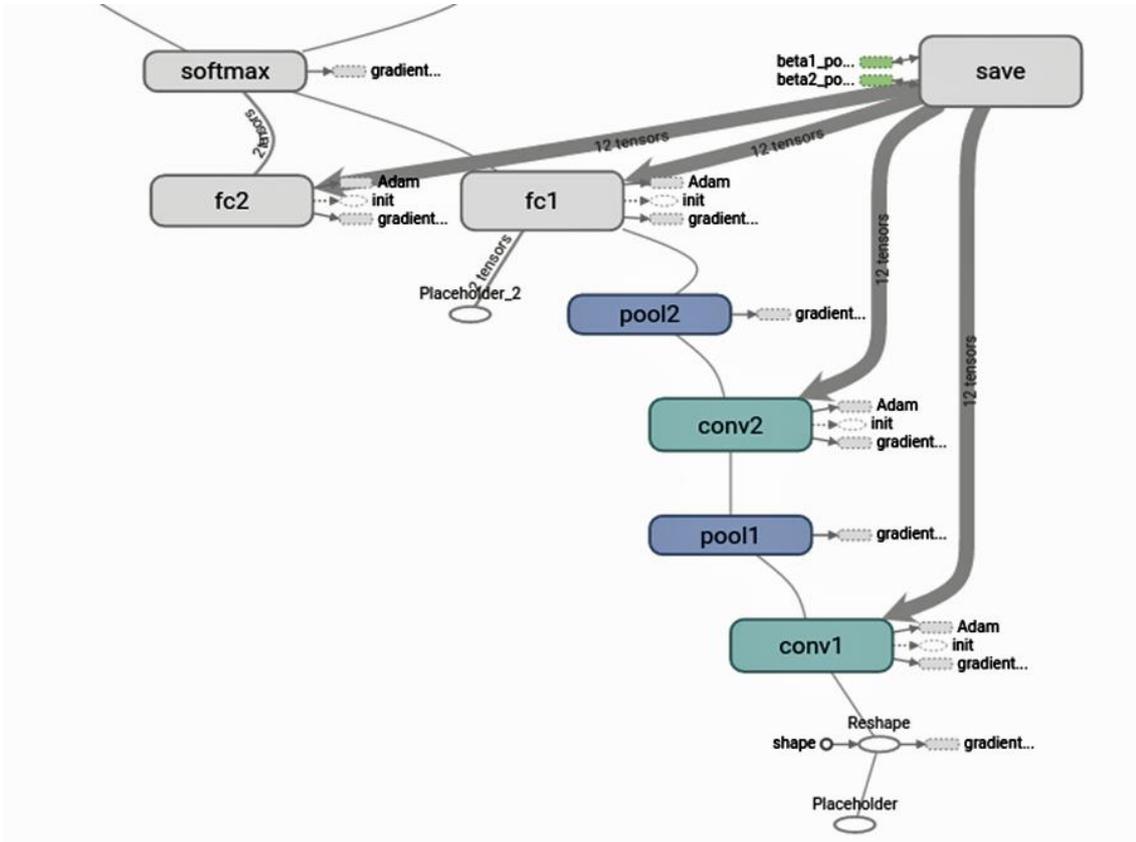

Fig. 4: Kernel part of final judgement code.

```python
if __name__ == '__main__':
    test_image = []
    for i in range(1, len(sys.argv)):
        img = cv2.imread(sys.argv[i])
        img = cv2.resize(img, (56, 56))
        test_image.append(img.flatten().astype(np.float32)/255.0)
    test_image = np.asarray(test_image)

    images_placeholder = tf.placeholder("float", shape=(None, IMAGE_PIXELS))
    labels_placeholder = tf.placeholder("float", shape=(None, NUM_CLASSES))
    keep_prob = tf.placeholder("float")

    logits = inference(images_placeholder, keep_prob)
    sess = tf.InteractiveSession()

    saver = tf.train.Saver()
    sess.run(tf.initialize_all_variables())
    saver.restore(sess, "model.ckpt")

    for i in range(len(test_image)):
        pred = logits.eval(feed_dict={
            images_placeholder: [test_image[i]],
            keep_prob: 1.0 })
        print pred
```

Fig. 5: Plots of step count against pixel size at training stage. Here, step count means the number of steps required to achieve the judgement accuracy of higher than 0.9. Blue and red lines correspond to settings of fixed dropout and cascade dropout (see text).

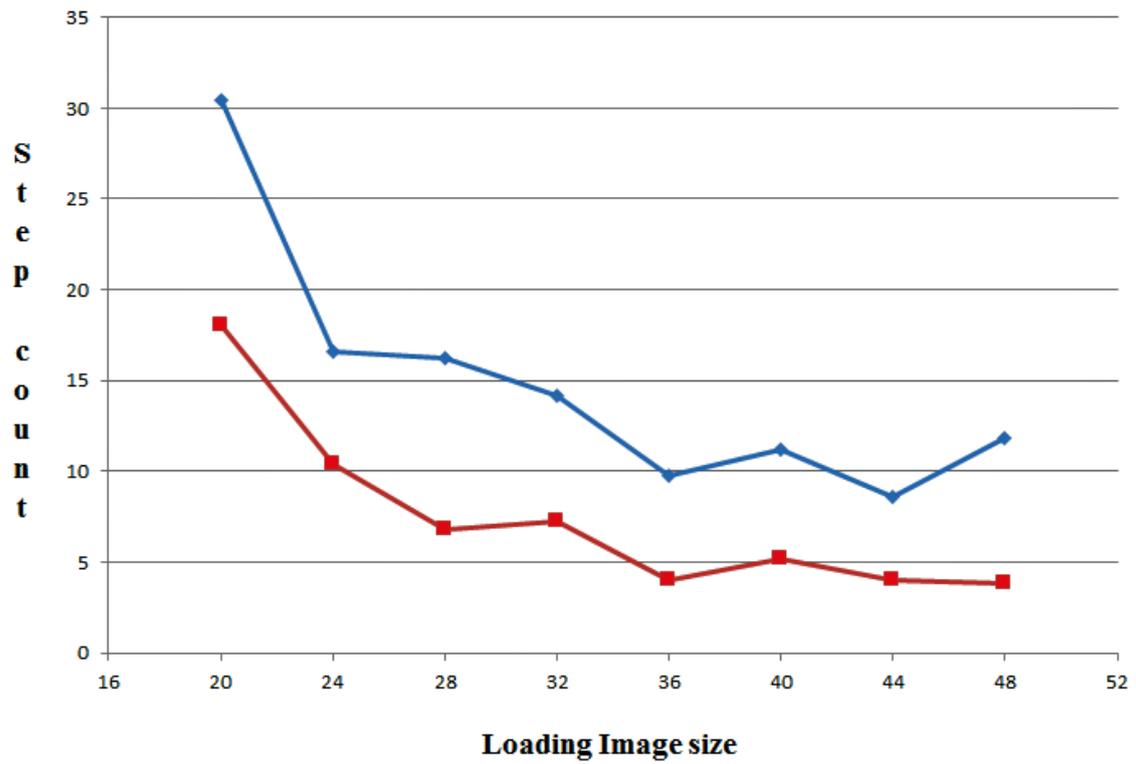